\documentclass[11pt]{article}

\usepackage{geometry}
\usepackage{currfile}
\usepackage{textcomp}

\usepackage{xcolor,soul}
\sethlcolor{yellow}

\usepackage[pdftex]{graphicx}
\DeclareGraphicsExtensions{.pdf,.jpeg,.png}
\graphicspath{{./}}

\date{May 15, 2019}

\begin{document}

\title{\Large{Data Cooperatives:\\
Towards a Foundation for Decentralized Personal Data Management}\\
~~}
\author{
\large{Thomas Hardjono and Alex Pentland}\\
\large{~~}\\
\large{MIT Connection Science}\\
\large{Massachusetts Institute of Technology}\\
\large{Cambridge, MA, USA}\\
\large{~~}\\
\small{{\tt hardjono@mit.edu}~~{\tt sandy@media.mit.edu}}\\
\large{~~}\\
}

\maketitle

\begin{abstract}
Data cooperatives with fiduciary obligations to members provide
a promising direction for the empowerment of individuals through their own personal data.
A data cooperative can manage, curate and protect access to the personal data of citizen members.
Furthermore, the data cooperative can run internal analytics in order to obtain insights regarding
the well-being of its members.
Armed with these insights, the data cooperative would be in a good position
to negotiate better services and discounts for its members.
Credit Unions and similar institutions 
can provide a suitable realization of data cooperatives.

~~\\
\end{abstract}

\newpage
\clearpage

%%%%%%%%%%%%%%%%	(NB. Table of Contents will be deleted in the final version)

%%%%%%%%%%%%%%%%	\par\noindent\rule{\textwidth}{0.4pt}

%%%%%%%%  \tableofcontents

%%%%%%%%  \par\noindent\rule{\textwidth}{0.4pt}

%%%%%%%%%%%%%%%%%%%%%%%%%%%%%%%%%%%%%%%%%%%%%%%%%%%%%%%%%%%%%
\section{Introduction}

During the last decade, all segments of society have become increasingly 
alarmed by the amount of data, and resulting power, held by a small number of actors~\cite{PentlandHardjono2019a}.  
Data is, by some, famously called ``the new oil''~\cite{WEF2011} and comes 
from records of the behavior of citizens. 
Why then, is control of this powerful new  resource concentrated in so few hands?   
During the last 150 years, questions about concentration of power 
have emerged each time the economy has shifted to a new paradigm; 
industrial employment replacing agricultural employment, consumer banking replacing cash and barter, 
and now ultra-efficient digital businesses 
replacing traditional physical businesses and civic systems.   

As the economy was transformed by industrialization and then by consumer banking, 
powerful new players such as Standard Oil, J.P. Morgan, and a handful of others 
threatened the freedom of citizens. 
In order to provide a counterweight to these new powers, 
citizens joined together to form trade unions and 
cooperative banking institutions, which were federally chartered 
to represent their members' interests. 
These citizen organizations helped balance 
the economic and social power between 
large and small players and between employers and worker.

The same collective organization is required to move 
from an individualized asset-based understanding of data control 
to a collective system based on rights and accountability, 
with legal standards upheld by a new class of representatives 
who act as fiduciaries for their members.  
In the U.S. almost 100 million people are members of credit unions, 
not-for-profit institutions owned by their members, 
and already chartered to securely manage their members? 
Digital data and to represent them in a wide variety of financial transactions, 
including insurance, investments, and benefits. 
The question then is, could we apply the same push for citizen power 
to the area of data rights in the ever-growing digital economy?   

Indeed, with advanced computing technologies it is practically possible 
to automatically record and organize all the data that 
citizens knowingly or unknowingly give to companies and the government, 
and to store these data in credit union vaults.
In addition, almost all credit unions already manage their 
accounts through regional associations that use common software, 
so widespread deployment of data cooperative capabilities 
could become surprisingly quick and easy.

%%%%%%%%%%%%%%%%%%%%%%%%%%%%%%%%%%%%%%%%%%%%%%%%%%%%%%%%%%%%%%%%%%%%%%%%%%%
\section{Data Cooperatives as Citizens' Organizations}
\label{sec:DataCoopIntro}

The notion of a {\em data cooperative} refers to the voluntary collaborative
pooling by individuals of their personal data
for the benefit of the membership of the group or community.
The motivation for individuals to get together and pool their data
is driven by the need to share common insights 
across data that would be otherwise siloed
or inaccesible.
These insights provide the cooperative members as a whole with a better understanding
of their current economic, health and social conditions
as compared to the other members of the cooperative generally.

A case in point would be the income made by ride-share automobile drivers (e.g. Uber, Lyft, etc). 
Currently, it is difficult for drivers to compare their respective 
incomes across similar routes, areas and distances. 
Similarly, passengers or riders who use these ride-share services 
have little or no idea whether their average fees for a given 
distance and city-sector is comparable to similar situations 
in a different sector of the city. 
By pooling together data accessible to them on their devices, 
drivers and passengers are able to see whether on average 
they are obtaining similar quality of service and equitable payment across a wide geographic area.

There are several key aspects to the notion of the data cooperative:
\begin{itemize}

\item	{\em Individual members own and control their personal data}:
The individual as a {\em member} of the data cooperative
has unambiguous legal ownership of (the copies of) their data.
Each member can collect copies of their data through various means,
either automatically using electronic means 
(e.g. passive data-traffic copying software on their devices) 
or by manually uploading data files to the cooperative.
This data is collected into the member's
{\em personal data store} (PDS)~\cite{openPDS2014PLOS}.
The member is able to add, subtract or remove data
from their personal data store, and even suspend access to their data store.
A member may posses multiple personal data repositories.

The member has the option to maintain their personal data store at the cooperative,
or host it elsewhere (e.g. private data server, cloud provider, etc).
In the case where the member chooses to host the personal data store at the cooperative,
the cooperative has the task of protecting the data
(e.g. encryption for data loss prevention)
and optionally curating the data sets for 
the benefit of the member (e.g. placing into common format, 
providing informative graphical reporting, etc.).

%%%~\cite{openPDS2014PLOS,Hardjono1996a,HardjonoSeberry97a}.

\item	{\em Fiduciary obligations to members}:
The data cooperative has a legal fiduciary obligation 
first and foremost to its members~\cite{Balkin2016}.
The organization is member-owned and member-run,
and it must be governed by rules (bylaws)
agreed to by all the members.

A key part of this governance rules
is to establish clear policies regarding the usage or access
to data belonging to its members.
These policies have direct influence on the work-flow of data access
within the cooperative's infrastructure,
which in turn has impact on how data privacy is enforced within the organization.

\item	{\em Direct benefit to members}:
The goal of the data cooperative is to benefit its members 
first and foremost.
The goal is not to ``monetize'' their data, 
but instead to perform on-going analytics
to understand better the needs of the members
and to share insights among the members.

\end{itemize}

There are numerous ways for a data cooperative to provide value to its members. 
For example, the cooperative could perform data analysis 
related to the health and age of its members, 
based for example on location data-sets. 
It may find, for instance, that a certain subset of the membership 
is not sufficiently paying attention to their health 
(e.g. not using available medical services). 
In such cases, the cooperative could devise strategies 
to remedy the situation, such as intervening and/or negotiating 
with external providers for better service rates (e.g. discounts to sporting facilities, health services, etc). 
Thus, these insights provide the cooperative with better bargaining power when it negotiates group purchases.

%%%%%%%%%%%%%%%%%%%%%%%%%%%%%%%%%%%%%%%%%%%%%%%%%%%%%%%%%%%%%%%%%%%%%%%%%%%
\section{The Data Cooperative Ecosystem}
\label{sec:DataCoopEcosystem}

\begin{figure}[!t]
\centering
\includegraphics[width=1.0\textwidth, trim={0.0cm 0.0cm 0.0cm 0.0cm}, clip]{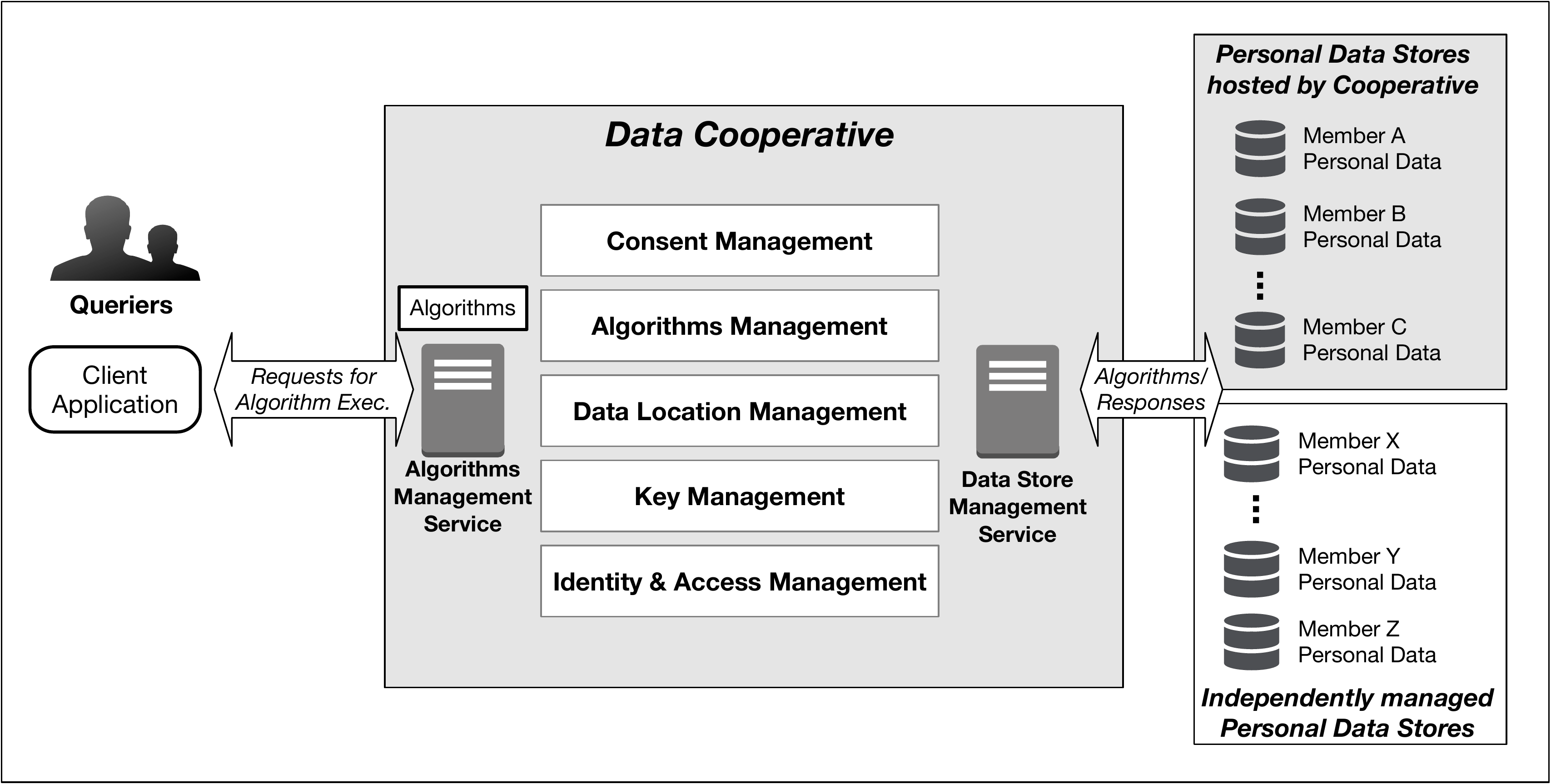}
	%
	% TRIMMING:  trim={<left> <lower> <right> <upper>} and clip options:
	% FULL EXAMPLE: \includegraphics[width=0.4\textwidth, trim={0.5cm 0.5cm 0.5cm 11.3cm}, clip]{image1.pdf}
	%
\caption{Overview of the Data Cooperative Ecosystem}
\label{fig:data-coop-ecosystem}
\end{figure}

The data cooperative ecosystem is summarized in Figure~\ref{fig:data-coop-ecosystem}.
The main entities are the (i) data cooperative as a legal entity,
(ii) the individuals who make-up the membership and elect the leadership of the cooperative,
and 
(iii) the external entities who interact with the data cooperative,
referred to as {\em queriers}.

The cooperative as an organization may choose to operate its own 
IT infrastructure or choose to outsource these IT functions to 
an external operator or IT services provider. 
In the case of outsourcing, the service level agreement (SLA) 
and contracts must include the prohibition for the operators 
to access or copy the members data. 
Furthermore, the prohibition must extend to all other third-party 
entities from which the outsourcing operator purchases or subcontracts parts if its own services.

A good analogy can be gleaned from Credit Unions throughout the United States.
Many of the small credit unions band together to share IT costs
by outsourcing IT services from a common provider,
known in industry as {\em Credit Union Service Organizations} (CUSO).
Thus, a credit union in Vermont may band together with
one in Texas and another in California,
to contract a CUSO to provide basic IT services.
This includes a common computing platform on the cloud,
shared storage on the cloud, shared applications, and so on.
The credit union may not have any equipment on-premises,
other than the PC computers used to connect 
to the platform operated by the CUSO.
Here, despite the three credit unions
using a common platform
the CUSO may tailor the appearance of the user interface differently for each
credit union in order to provide some degree of differentiation
to its members.
However, the CUSO in turn may be subcontracting 
functions or applications from a third party.
For example, the CUSO may be running its platform
using virtualization technology on Amazon Web Services (AWS).
It may purchase storage from yet a different entity.
This approach of subcontracting functions or services
from other service provider is currently very common.

In the context of the data cooperative that choses to outsource IT services,
the service contract with the IT services provider
must include prohibitions by third party
cloud providers from accessing data belonging to
the cooperative's members.

%%%%%%%%%%%%%%%%%%%%%%%%%%%%%%%%%%%%%%%%%%%%%%%%%%%%%%%%%%%%%%%%%%%%%%%%%%%
\section{Preserving Data Privacy of Members}
\label{sec:Dataprivacy}

We propose to use the MIT {\em Open Algorithms} (OPAL) approach to ensure the privacy
of the member's data held within the personal data stores.
In essence, the OPAL paradigm requires that data never be moved or 
be copied out of its data store,
and that the algorithms are instead transmitted to the data stores for execution.

The following are the key concepts and principles underlying the 
open algorithms paradigm~\cite{PentlandShrier2016}:
\begin{itemize}

\item	{\em Move the algorithm to the data}: Instead of ``pulling'' data into
a centralized location for processing, it is the algorithm that 
must be transmitted to the data repositories endpoints and be processed there.

\item	{\em Data must never leave its repository}: Data must never be exported
or copied from its repository.
Additional local data-loss protection could be applied, such as encryption
(e.g. homomorphic encryption) to prevent backdoor theft of the data.

\item	{\em Vetted algorithms}:  Algorithms must be vetted 
to be ``safe'' from bias, discrimination, privacy violations and other unintended consequences.

\item	{\em Provide only safe answers}: When returning results
from executing one or more algorithms,
return  {\em aggregate answers} only as the default granularity of the response.

\end{itemize}
Aggregate responses must be granular enough so that it does not 
allow the recipient (e.g. querier entity) to perform correlation 
attacks that re-identify individuals. 
Any algorithm that is intended to yield answers that are 
specific to a data subject (individual) must only be 
executed after obtaining the subject?s affirmative and fully informed consent~\cite{GDPR}.

%%%%%%%%%%%%%%%%%%%%%%%%%%%%%%%%%%%%%%%%%%%%%%%%%%%%%%%%%%%%%%%%%%%%%%%%%%%
\section{Consent for Algorithm Execution}
\label{sec:Consent}

One of the contributions of the EU GDPR regulation~\cite{GDPR}
is the formal recognition at the regulatory level for
the need for {\em informed consent} to be obtain from subjects.
More specifically, the GDPR calls
for the ability for the entity processing the data
to
\begin{quote}
{\em ...demonstrate that the data subject has consented 
to processing of his or her personal data} (Article~7).
\end{quote}
Related to this, a given
\begin{quote}
{\em ...data subject shall have the right to 
withdraw his or her consent at any time} (Article~7).
\end{quote}
In terms of minimizing the practice of copying data unnecessarily,
the GDPR calls out in clear terms the need
to access data 
\begin{quote}
{\em ...limited to what is necessary in 
relation to the purposes for which they are processed (data minimisation)} (Article~5).
\end{quote}
In the context of the GDPR, 
we believe that the MIT Open Algorithms approach 
substantially addresses the various issues raised by the GDPR
by virtue of data never being moved or copied from its repository.

Furthermore,
because OPAL requires algorithms to be selected and transmitted
to the data endpoints for execution,
the matter of consent in OPAL becomes
one of {\em requesting permission from the subject 
for the execution of one or more vetted algorithms on the subject's data}.
The data cooperative as a member organization
has the task of explaining in lay terms
the meaning and purpose of each algorithm,
and convey to the members the benefits from executing
the algorithm on the member's data.

In terms of the consent management system implementation by a data cooperative,
there are additional requirements that pertain to {\em indirect access}
by service providers and operators that may be 
hosting data belonging to members of the cooperative.
More specifically,
when an entity employs a third-party operated
service (e.g. client or application running in the cloud)
and that service handles data, algorithms and computation results
related to the cooperative's activities,
then we believe authorization must be expressly obtained by that third-party.

\begin{figure}[!t]
\centering
\includegraphics[width=0.8\textwidth, trim={0.0cm 0.0cm 0.0cm 0.0cm}, clip]{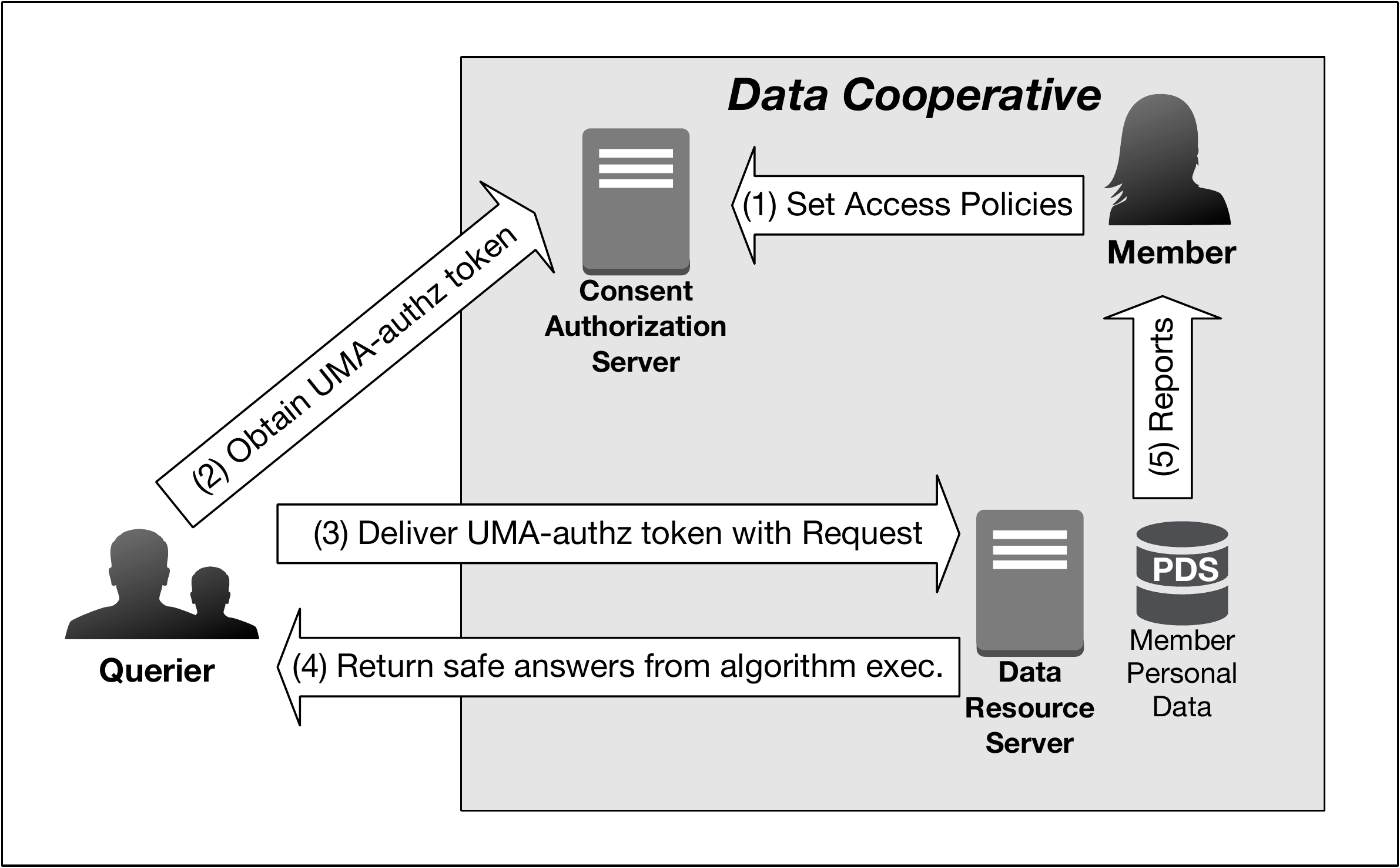}
	%
	% TRIMMING:  trim={<left> <lower> <right> <upper>} and clip options:
	% FULL EXAMPLE: \includegraphics[width=0.4\textwidth, trim={0.5cm 0.5cm 0.5cm 11.3cm}, clip]{image1.pdf}
	%
\caption{Consent Management using User Managed Access (UMA)}
\label{fig:UMA}
\end{figure}

In the context of possible implementations of authorization
and consent management,
the current popular access authorization framework
used by most hosted application and services providers today
is based on the {\em OAuth2.0} authorization framework~\cite{rfc6749}.
The {OAuth2.0} model is relatively simple in that it recognizes three (3)
basic entities in the authorization work-flow.
The first entity is the {\em resource-owner},
which in our case translates to the cooperative on behalf of its members.
The second entity is the {\em authorization service},
which could map to either the cooperative or an outsourced provider.
The third entity is the requesting party using a {\em client} (application),
which maps roughly to our querier (person or organization seeking insights).
In the case that the data cooperative is performing internal analytics for its own purposes,
then the querier is the cooperative itself.

While the {OAuth2.0} model has gained traction in industry over the past decade (e.g. in mobile apps),
its simplistic view of the 3-party world does not take
into account the reality today of the popularity of hosted applications and services.
In reality the three parties in {OAuth2.0} (namely the client, the authorization server and the resource)
could each be operated by separate legal entities.
For example, the client application could be running in the cloud,
and thus any information or data passing through the client application
becomes accessible to the cloud provider.

\begin{figure}[!t]
\centering
\includegraphics[width=0.8\textwidth, trim={0.0cm 0.0cm 0.0cm 0.0cm}, clip]{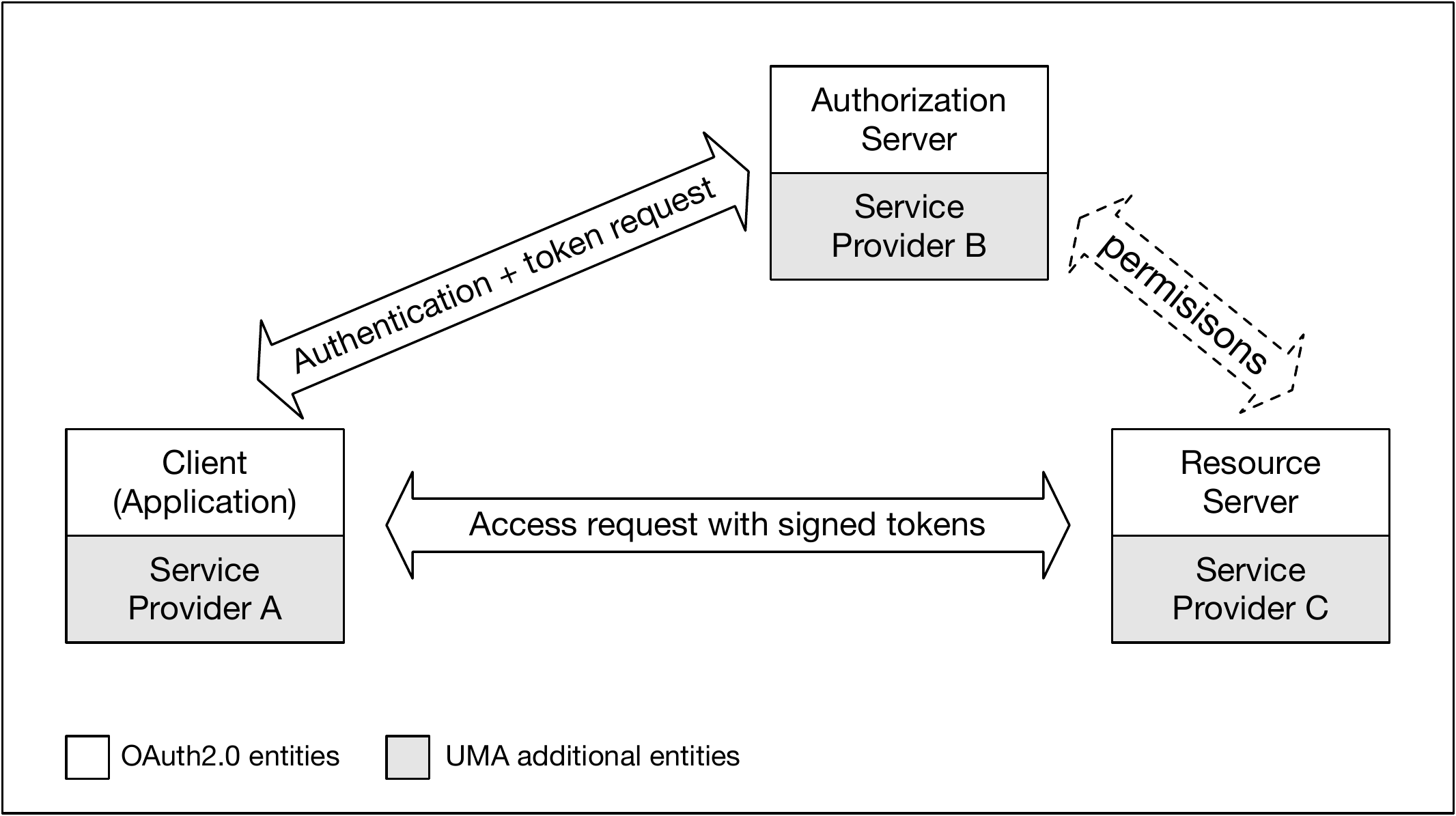}
	%
	% TRIMMING:  trim={<left> <lower> <right> <upper>} and clip options:
	% FULL EXAMPLE: \includegraphics[width=0.4\textwidth, trim={0.5cm 0.5cm 0.5cm 11.3cm}, clip]{image1.pdf}
	%
\caption{UMA entities as an extension of the {OAuth2.0} model}
\label{fig:UMAentities}
\end{figure}

An early awareness of the inherent limitations of {OAuth2.0}
has led to additional efforts to be directed at expanding
the 3-party configuration to become a 5 or 6 party arrangement (Figure~\ref{fig:UMAentities}),
while retaining the same {OAuth2.0} token and messaging formats.
This work has been conducted in the Kantara Initiative standards organization since 2009,
under the umbrella of {\em User Managed Access} (UMA)~\cite{UMACORE1.0,UMACORE2.0}.
As implied by its name, UMA seeks to provide ordinary users
as resource (data) owners
with the ability to manage access policy in a consistent
manner across the user's resources that maybe physically distributed
throughout different repositories on the Internet.
The UMA entities follows closely and extends the entities defined
in the {OAuth2.0} framework.
More importantly, the UMA model introduces new functions and tokens
that allow it to address complex scenarios that explicitly identity
hosted services providers and cloud operators as entities
that must abide by the same consent terms of service:
\begin{itemize}

\item	{\em Recognition of service operators as 3rd party legal entities}:
The UMA architecture explicitly calls-out entities which provide services
to the basic {OAuth2.0} entities.
The goal is to extend the legal obligations to these entities as well,
which is crucial for implementing informed consent in the sense of the GDPR.

Thus, for example, in the UMA work-flow in Figure~\ref{fig:UMAentities},
the client is recognized to be consisting of two separate
entities:  the querier (e.g. person) that that operates the hosted client-application,
and the Service Provider~A that makes available the client-application
on its infrastructure.
When the querier is authenticated by the authorization server and is issued
an access-token,
the Service Provider~A 
must also be separately authenticated and be issued a unique access token. 

This means that Service Provider~A which operates the client-application
must accept the terms of service and data usage agreement
presented by the authorization server,
in the same manner that the querier (person or organization) must accept them.

\item	{\em Multi-round handshake as a progressive legal binding mechanism}:
Another important contribution of the UMA architecture
is the recognition that a given endpoint (e.g. API at the authorization server)
provides the opportunity to successively engage the caller
to agree to a terms of service and data usage agreement (referred to as {\em binding obligations} in UMA).

More specifically,
UMA uses the multi-round protocol run between the client and the authorization server
to {\em progressively bind} the client in a lock-step manner.
When the client (client-operator)
chooses to proceed with the handshake by sending
the next message in the protocol to the endpoint of the authorization server,
the client has implicitly agreed to the terms of service at that endpoint.
This is akin to the client agreeing step-by-step to additional clauses in a contract
each time the client proceeds with the next stage of the handshake.

\end{itemize}

%%%%%%%%%%%%%%%%%%%%%%%%%%%%%%%%%%%%%%%%%%%%%%%%%%%%%%%%%%%%%%%%%%%%%%%%%%%
\section{Identity-related Algorithmic Assertions}
\label{sec:IdentityAssertion}

A potential role for a data cooperative is to make available the summary results
of analytic computations to external entities regarding a member (subject)
upon request by the member.
Here the work-flow must be initiated by the member who is using his or her data
(in their personal data store) as the basis for generating the assertions about them,
based on executing one or more of the cooperative-vetted algorithms.
In this case, the cooperative behaves as an {\em Attribute Provider}
or {\em Assertions Provider} for its members~\cite{ABA2012},
by issuing a signed assertion in a standard format (e.g. SAML2.0~\cite{SAMLcore} or Claims~\cite{Sporny2019,W3C-DID-2018}).
This is particularly useful when the member
is seeking to obtains goods and services from 
an external {\em Service Provider} (SP).

As an example,
a particular member (individual) could be seeking a loan (e.g. car loan)
from a financial institution.
The financial institution requires proof of incomes and expenditures
regarding the member over a duration of time (e.g. last 5 years),
as part of its risk assessment process.
It needs an authoritative and truthful source of information regarding
the member's financial behavior over the last 5 years.
This role today in the United States is fulfilled by the so called
credit scoring or credit report companies,
such as Equifax, TransUnion and Experian.

However, in this case the member could turn to its cooperative
and request the cooperative to run various algorithms 
-- including algorithms private to the cooperative -- 
on the various data sets regarding the member located in the member's personal data store.
At the end of these computations the cooperative could issue
an authoritative and truthful assertion,
which it signs using its private-key.
The digital signature signifies that the cooperative 
stands behind its assertions regarding the given member.
Then the cooperative or the member could transmit the
signed assertion to the financial institution.
Note that this cycle of executing algorithms,
followed by assertions creation and transmittal
to the financial institution can be repeated
as many times as needed,
until the financial institution is satisfied.

\begin{figure}[!t]
\centering
\includegraphics[width=0.8\textwidth, trim={0.0cm 0.0cm 0.0cm 0.0cm}, clip]{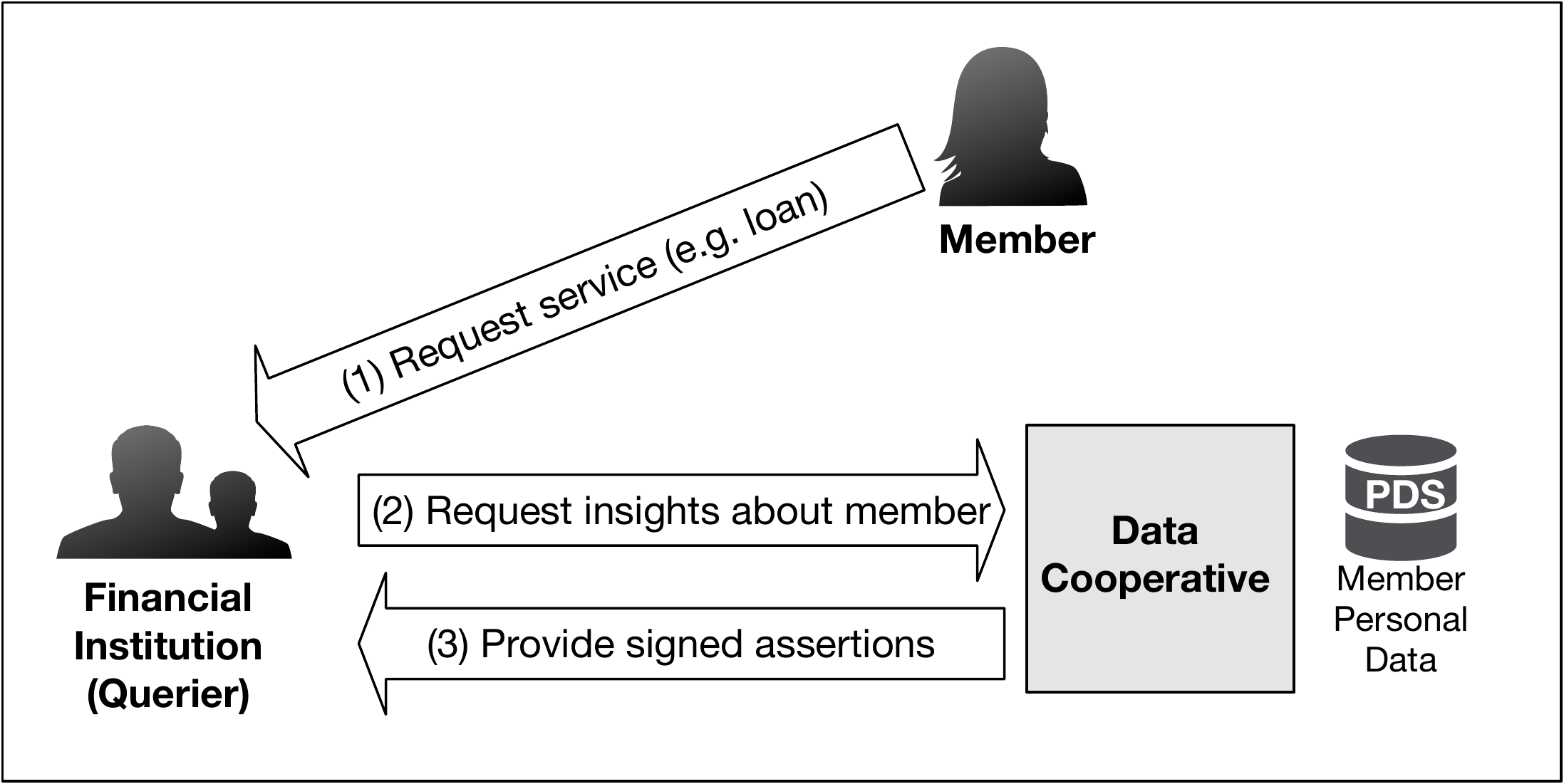}
	%
	% TRIMMING:  trim={<left> <lower> <right> <upper>} and clip options:
	% FULL EXAMPLE: \includegraphics[width=0.4\textwidth, trim={0.5cm 0.5cm 0.5cm 11.3cm}, clip]{image1.pdf}
	%
\caption{Overview of Obtaining Assertions from the Data Cooperative}
\label{fig:Assertions}
\end{figure}

There are a number of important aspects regarding this approach
of relying on the data cooperative:
\begin{itemize}

\item	{\em Member driven}: 
The algorithmic computation on the member's data
and the assertion issuance must be invoked or initiated by the member.
The data cooperative must not perform this computation and issue assertions (about a member)
without express directive from the member.

\item	{\em Short-lived assertions}: 
The assertion validity period should be very
limited to the duration of time specified by the service provider.
This reduces the window of opportunity
for the service provider to hoard
or re-sell to a third party the assertions obtained from the cooperative.

\item	{\em Limited to a specific purpose}:  
The assertion should carry additional legal clause
indicating that the assertion is to be used for a given purpose only
(e.g. member's application for a specific loan type and loan amount).

\item	{\em Signature of cooperative}: The data cooperative
as the issuer of the assertions or claims
must digitally sign the assertions.
This conveys the consent of the member (for the issuance of assertion)
and conveys the authority of the cooperative as the entity
who executes algorithms over the member's data.

\item	{\em Portability of assertions}: The assertion data structure 
should be independent (stand-alone),
portable and not tied to any specific infrastructure.

\item	{\em Incorporates Terms of Use}: 
The assertion container (e.g. {SAML2.0} or Claims)
issued by the cooperative
must carry unambiguous legal statements regarding 
the terms of use of the information
contained in the assertion.
The container itself may even carry a copyright notice
from the cooperative to discourage service providers from
propagating the signed assertions to third parties.

\end{itemize}

Once the assertion has been issued by the cooperative
there are numerous ways to make the assertion
available to external third parties -- depending
on the privacy limitations of the concerned entities.
In the case above,
a member (subject) may wish for the assertion to be
availably only to the specific service provider (e.g. loan provider)
because the event pertains to a private transaction.
In the case that the service provider needs to maintain copies
of assertions from the cooperative 
for legal reasons (e.g. taxation purposes),
the service provider could return 
a signed digital receipt~\cite{CONSENT1.0}
agreeing to the terms of use of the assertions.

In other cases, a member may wish for some types of assertions
containing static personal attributes (e.g. age or year of birth)
to be readily available without the privacy limitations.
For example, the member might use such attribute-based assertions
to purchase merchandise tied to age limits (e.g. alcohol).
In this case, the signed assertion can be readable from a well-known
endpoint at the cooperative, 
be readable for the member's personal website,
or be carried inside the member's mobile device. 
Hence the importance of the portability of the assertions structure.

%%%%%%%%%%%%%%%%%%%%%%%%%%%%%%%%%%%%%%%%%%%%%%%%%%%%%%%%%%%%%%%%%%%%%%%%%%%
\section{Conclusions}
\label{sec:Conclusions}

Today we are in a situation where individual assets ...people's personal data... is being 
exploited without sufficient value being returned to the individual.  
This is analogous to the situation in the late 1800's and early 1900's 
that led to the creation of collective institutions 
such as credit unions and labor unions, 
and so the time seems ripe for the creation of 
collective institutions to represent the data rights of individuals. 

We have argued that data cooperatives with fiduciary obligations 
to members provide a promising direction for the empowerment 
of individuals through collective use of their own personal data. 
Not only can a data cooperative give the individual expert, 
community-based advice on how to manage, curate and protect 
access to their personal data, it can run internal analytics that 
benefit the collective membership. 
Such collective insights provide a powerful tool for negotiating 
better services and discounts for its members. 
Federally chartered Credit Unions are already legally 
empowered to act as data cooperatives, and we believe 
that there are many other similar institutions that could also provide data cooperative services.

%%% \bibliographystyle{IEEEtran}
%%% \bibliography{IEEEabrv,hardjonobib,thomasrfcbib}

\begin{thebibliography}{10}
\providecommand{\url}[1]{#1}
\csname url@samestyle\endcsname
\providecommand{\newblock}{\relax}
\providecommand{\bibinfo}[2]{#2}
\providecommand{\BIBentrySTDinterwordspacing}{\spaceskip=0pt\relax}
\providecommand{\BIBentryALTinterwordstretchfactor}{4}
\providecommand{\BIBentryALTinterwordspacing}{\spaceskip=\fontdimen2\font plus
\BIBentryALTinterwordstretchfactor\fontdimen3\font minus
  \fontdimen4\font\relax}
\providecommand{\BIBforeignlanguage}[2]{{%
\expandafter\ifx\csname l@#1\endcsname\relax
\typeout{** WARNING: IEEEtran.bst: No hyphenation pattern has been}%
\typeout{** loaded for the language `#1'. Using the pattern for}%
\typeout{** the default language instead.}%
\else
\language=\csname l@#1\endcsname
\fi
#2}}
\providecommand{\BIBdecl}{\relax}
\BIBdecl

\bibitem{PentlandHardjono2019a}
A.~Pentland, T.~Hardjono, J.~Penn, and C.~Colclough, ``{D}ata {C}ooperatives:
  {D}igital {E}mpowerment of {C}itizens and {W}orkers,'' January 2019.

\bibitem{WEF2011}
{World Economic Forum}, ``{P}ersonal {D}ata: {T}he {E}mergence of a {N}ew
  {A}sset {C}lass,'' 2011, http://www.weforum.org/reports/
  personal-data-emergence-new-asset-class.

\bibitem{openPDS2014PLOS}
Y.~A. {de~Montjoye}, E.~Shmueli, S.~Wang, and A.~Pentland, ``{openPDS}:
  {P}rotecting the {P}rivacy of {M}etadata through {SafeAnswers},'' \emph{PLoS
  ONE 9(7)}, pp. 13--18, July 2014,
  https://doi.org/10.1371/journal.pone.0098790.

\bibitem{Balkin2016}
J.~M. Balkin, ``{I}nformation {F}iduciaries and the {F}irst {A}mendment,''
  \emph{UC Davis Law Review}, vol.~49, no.~4, pp. 1183--1234, April 2016.

\bibitem{PentlandShrier2016}
A.~Pentland, D.~Shrier, T.~Hardjono, and I.~{Wladawsky-Berger}, ``{T}owards an
  {I}nternet of {T}rusted {D}ata: {I}nput to the {W}hitehouse {C}ommission on
  {E}nhancing {N}ational {C}ybersecurity,'' in \emph{{Trust::Data} - {A} {N}ew
  {F}ramework for {I}dentity and {D}ata {S}haring}, T.~Hardjono, A.~Pentland,
  and D.~Shrier, Eds.\hskip 1em plus 0.5em minus 0.4em\relax Visionary Future,
  2016, pp. 21--49.

\bibitem{GDPR}
{European Commission}, ``Regulation {(EU)} 2016/679 of the {E}uropean
  {P}arliament and of the {C}ouncil of 27 {A}pril 2016 on the protection of
  natural persons with regard to the processing of personal data and on the
  free movement of such data ({G}eneral {D}ata {P}rotection {R}egulation),''
  \emph{Official Journal of the European Union}, vol. L119, pp. 1--88, 2016.

\bibitem{rfc6749}
\BIBentryALTinterwordspacing
D.~Hardt, ``{T}he {OAuth~2.0} {A}uthorization {F}ramework,'' October 2012,
  {RFC6749}. [Online]. Available: \url{http://tools.ietf.org/rfc/rfc6749.txt}
\BIBentrySTDinterwordspacing

\bibitem{UMACORE1.0}
T.~Hardjono, E.~Maler, M.~Machulak, and D.~Catalano, ``{U}ser-{M}anaged
  {A}ccess ({UMA}) {P}rofile of {OAuth2.0} -- {S}pecification {V}ersion
  {1.0},'' Kantara Initiative, Kantara Published Specification, April 2015,
  https://docs.kantarainitiative.org/uma/rec-uma-core.html.

\bibitem{UMACORE2.0}
E.~Maler, M.~Machulak, and J.~Richer, ``{U}ser-{M}anaged {A}ccess ({UMA})
  {2.0},'' Kantara Initiative, Kantara Published Specification, January 2017,
  https://docs.kantarainitiative.org/uma/ed/uma-core-2.0-10.html.

\bibitem{ABA2012}
{American Bar Association}, ``{A}n {O}verview of {I}dentity {M}anagement:
  {S}ubmission for {UNCITRAL} {C}ommission {45th} {S}ession,'' {ABA Identity
  Management Legal Task Force}, May 2012, available on
  http://meetings.abanet.org/ webupload/commupload/ CL320041/relatedresources/
  ABA-Submission-to-UNCITRAL.pdf.

\bibitem{SAMLcore}
{OASIS}, ``{A}ssertions and {P}rotocols for the {OASIS} {S}ecurity {A}ssertion
  {M}arkup {L}anguage ({SAML}) {V2.0},'' March 2005, available on
  {http://docs.oasisopen.org/security/ saml/v2.0/ saml-core-2.0-os.pdf}.

\bibitem{Sporny2019}
M.~Sporny, D.~Longley, and D.~Chadwick, ``{V}erifiable {C}redentials {D}ata
  {M}odel {1.0},'' {W3C}, {W3C} {C}andidate {R}ecommendation, March 2019,
  available at {https://www.w3.org/TR/verifiable-claims-data-model}.

\bibitem{W3C-DID-2018}
D.~Reed and M.~Sporny, ``{D}ecentralized {I}dentifiers ({DIDs}) {v0.11},'' W3C,
  Draft Community Group Report 09 July 2018, July 2018,
  https://w3c-ccg.github.io/did-spec/.

\bibitem{CONSENT1.0}
M.~Lizar and D.~Turner, ``{C}onsent {R}eceipt {S}pecification {V}ersion
  {1.0},'' March 2017,
  https://kantarainitiative.org/confluence/display/infosharing/Home.

\end{thebibliography}


\end{document}